# Un hecho estilizado en la revocación de mandato en México 2022 que recuerda las presidenciales de 2006

Hugo Hernández-Saldaña
[1]Departamento de Ciencias Básicas, UAM en Azcapotzalco, CdMx., México
Teléfono (55) 5318-9575     E-mail: hhs@azc.uam.mx

*Resumen* — **Las votaciones son el resultado de varias interacciones complejas y que resulta difícil de separar para entender el papel que juegan en el resultado final. El ejercicio revocatorio llevado a cabo en México en el año 2022 permite una aproximación acotada a la dinámica de voto en México. Pues, en principio, es binaria y la votación fue fundamentalmente debida al partido en el gobierno y a la coerción del voto, siendo estas las principales dinámicas de voto. En éste trabajo reportamos que la distribución de voto por casilla para el NO es parecida a la distribución que ha aparecido en diversas elecciones en votos para el Partido Revolucionario Institucional y es compatible con una distribución Gamma. Se discuten sus similitudes y dos posibles modelos para tal comportamiento, uno basado en el Problema del Agente Viajero (TSP, por sus sigla en inglés) y otro en una ecuación de Fokker-Planck.**

*Palabras Clave* – **distribución de voto, hechos estilizados, ,proceso electoral**

*Abstract* — **Vote processes are the results of several complex interactions which are hard to separate in order to understand its role in the final result. The revocatory exercise performed in Mexico in 2022 allows an approximation to the Mexican dynamics of vote. Being a binary election (in principle) and the vote was mainly due to the political party in the government and the coercion of vote we have mainly only two dynamics of vote. In this work I report that the vote distribution for the NO at urn scale is similar to the one that appears in several elections for the Institutional Revolutionary Party (PRI for his Spanish acronym) and it is compatible with a Gamma distribution. We discuss similarities and two possible models for such a behavior, one based on the Traveling Salesperson Problem and other on a Fokker-Planck equation.**

## I. Introducción

La búsqueda de leyes o patrones en la forma en la que las personas votan o actúan es un tema de investigación con un florecimiento en las últimas décadas [1-5]. En particular el uso de modelos importados de la mecánica estadística aplicados al creciente número de datos disponibles ha permitido algunos avances. La complejidad de los fenómenos sociales es suficiente para que la investigación sea muy activa. En el caso de los estudios enfocados al análisis de elecciones, un problema consiste en las diferentes opciones e interacciones que concurren en una elección dada, adicionado al proceso mismo de conteo del voto.

El ejercicio de revocación de mandato realizado en México el 10 de abril del año 2022 ofrece algunas oportunidades interesantes. Al ser promovida principalmente por el partido en el gobierno federal y su partido, el MORENA, se convirtió en una posible radiografía de dicho grupo político haciendo uso amplio de recursos económicos y políticos para la promoción del voto. Más aún, de acuerdo a una amplia evidencia [6], hubo coerción del voto y uso inadecuado de recursos públicos [7]. Este tipo de situaciones no son nuevas en México, donde un partido reguló la vida política por varias décadas haciendo uso de esas mismas estrategias electorales[8]. Dicho partido, el Partido Revolucionario Institucional, PRI, ha tenido una distribución de voto bastante regular a lo largo de las elecciones de las cuales tenemos datos a nivel casilla desde que existió una institución no gubernamental para hacer el conteo, a saber, el Instituto Federal Electoral, IFE. La distribución de voto del PRI en muchas de las contiendas federales en las que ha participado están bien descritas por una distribución Gamma en general, o una distribución de Margarita en particular, caracterizada por un decaimiento exponencial y un arranque en ley de potencia[9].

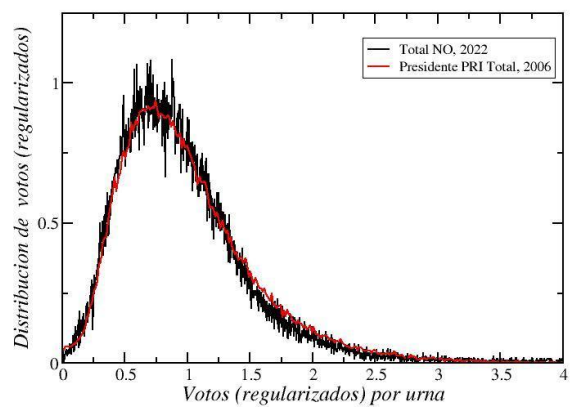

Fig. 1. Distribución de voto por urna por el NO en la elección de revocación de mandato 2022 (línea negra)y por el PRI en la elección de 2006 (línea roja). Se presenta el voto regularizado por su promedio, es decir, voto por casilla entre el promedio de votos.

Lo interesante es que, cuando uno hace la distribución de voto por casilla para el ejercicio revocatorio con la respuesta NO, su forma es compatible con una distribución del PRI para presidente en 2006 (véase la Fig. 1), por ejemplo. En dicha Fig. se compara la distribución de votos rectificados por el número de votos promedio para cada caso. Sobre cómo obtenemos la distribución de la Fig., su correspondencia con una posible dinámica de voto y algunas especificidades en la manera en que el voto se da a escala municipal será el tema del presente artículo.

## II. Metodología

Se hace uso de las bases de datos proporcionadas por las autoridades electorales federales a través de su página [10,11]. Dado que los recortes presupuestarios obligaron a disminuir el número de casillas y su distribución geográfica se verifica el histograma de número de votantes registrados para el ejercicio o lista nominal. Cómo se mostró en [12] el algoritmo de distritación agrupa el número de votantes máximo, la lista nominal, en al menos tres grupos claramente diferenciados [13]. En el presente ejercicio la decimación borra buena parte de los grupos como puede observarse en la Fig. 2. Ahí simplemente se realiza el histograma de la lista nominal, es decir cuántas casillas permiten 1 votantes, cuántas 2, etc. La mayoría se agrupa entre, aproximadamente, 1350 y 1950 votantes permitidos o simplemente la lista nominal. Ambos valores difieren un poco de haber compactado la lista nominal del ejercicio de 2021. Ahí la lista nominal tenía una zona con una gran cantidad de casillas con entre 550 y 750 votantes, dando 1650 y 2250 como los límites. Las otras zonas de casillas corresponden a un decaimiento para lista nominal mayor a 1950 y una meseta con valores entre mil y 1300, aproximadamente. En una estudio posterior se discutirá la decimación con parámetros distritales, seccionales y municipales. No se consideraron las casillas especiales ni el voto en el extranjero.

Aún cuando la pregunta en la boleta no era binaria, aquí la denominaremos así: SI y NO. El voto mayoritario fue

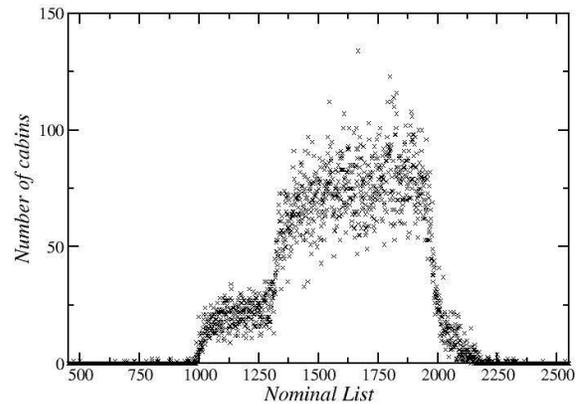

Fig. 2. Histograma de la lista nominal, o número máximo de electores habilitados en la casilla o urna, para el ejercicio de revocación de mandato en México 2022. No se realiza promediamiento ni suavización alguna. Magnetización en función del campo aplicado.

para el NO. Aquí lo denotaremos como NO-2022 o simplemente NO.

La distribución de voto por casilla total se contabiliza directamente, solo quitando las casillas que fueron anuladas, las especiales y el voto en el extranjero. Un estudio
detallado con consideraciones geográficas, históricas y su referente a factores de marginación está en progreso [14].

Aún cuando es usual el análisis de distribuciones sobre las variables estandarizadas aquí lo hacemos sobre la variable rectificada por su valor promedio, lo que nos da una distribución de promedio unitario. Por desgracia no hay una clara manera de jerarquizar los datos para analizar correlaciones de mediano y largo alcance.

## III. Resultados

### A. Descripción de las distribuciones

Como puede observarse en la Fig. 1 y repetimos en la Fig. 3, la distribución del NO claramente sigue una forma parecida a la del PRI en la elección presidencial del 2006. Para poder comparar las dos distribuciones de voto rectificamos el voto por casilla por su promedio y rehacemos el histograma. Aquí no presentamos los histogramas de voto simple. Las distribuciones que se obtienen después de la rectificación presentan un inicio como ley de potencia y un decaimiento exponencial. Esta

forma asimétrica tiene un buen ajuste con una distribución Gamma

$$P_{Gamma}(s) = N_{Gamma} s^\alpha Exp[-\beta s]. \quad (1)$$

Siendo $N_{Gamma}$ la constante de normalización. En esta distribución los parámetros estan elegidos de tal manera que el promedio sea uno. En la Fig. 3 mostramos los mismos resultados de voto y adicionamos los ajustes a una distribución Gamma para cada uno. Además se presenta el ajuste de una log-Normal que explicaremos más adelante. Esta distribución tiene la forma

$$P_{Log}(s) = N_{Log} x^\eta Exp[-\theta(log(s)-\gamma)^2]. \quad (2)$$

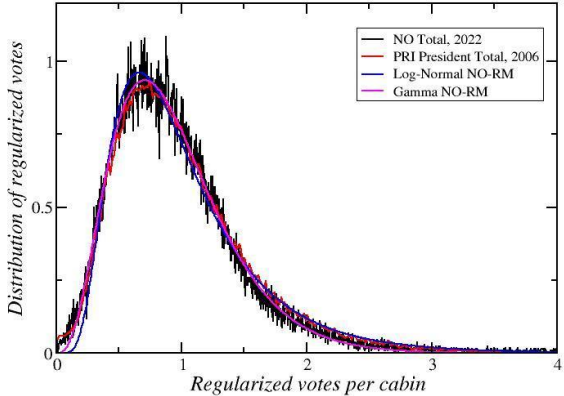

Fig. 3. Mismos histogramas que en la Fig. 2. Se han añadido los ajustes a los datos de revocación de mandato usando una distribución Log-Normal (línea azul) y una distribución Gamma (línea magenta).

De nuevo, los parámetros son elegidos para tener promedio unidad.

El PRI ha mostrado un buen ajuste a estas distribuciones en varias elecciones [15] aunque un análisis más cuidadoso muestra que el resultado puede generalizarse al reconsiderar la distribuciones de acuerdo al número de votantes máximos que permite hacer un estudio estadístico con muestreos adecuados [12].

En esta nota lo que deseamos resaltar es que el resultado presidencial 2006 del PRI correspondió a la votación de su núcleo duro pues con meses de anticipación las encuestas marcaban que su candidato iba en tercer lugar. Además el análisis de errores en el cómputo de los parámetros muestra el mejor ajuste a una curva suave, en este caso aquella dada por (1). Los otros partidos y los votos nulos muestran distribuciones diferentes en general. Así, la pregunta obvia es si tales distribuciones, compatibles con distribuciones Gamma, corresponden al voto duro de partidos corporativos. En el caso del ejercicio revocatorio es claro que el voto corresponde a dos vertientes principales: 1) el voto de los partidarios del partido en el gobierno y 2) los votantes coaccionados para votar. Y esa es la importancia de analizar con mayor detalle este ejercicio, permitiría entender el voto sin los agregados de votantes que simpatizan temporalmente con el partido/candidato en una elección particular.

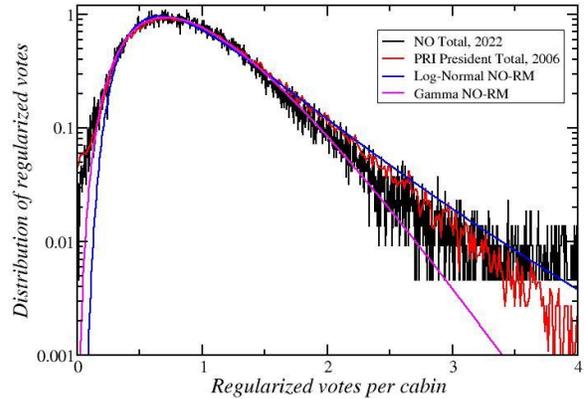

Fig. 4. Mismos histogramas que en la Fig. 3 pero en escala semilogarítmica. Note que el decaimiento del NO y el PRI-2006 es paralelo, no así los ajustes de las funciones (1) y (2).

Además de la discusión de los modelos que se hará en la sección siguiente una cosa interesante es que las distribuciones de voto presentan un problema de colas anchas, es decir, la distribución experimental decae más lento que la teórica, como puede apreciarse en la Fig. 4, donde se grafican las curvas de la Fig. 3 en escala semilogarítmica. Una distribución que es usada en el contexto de problemas de colas anchas es la de una log-normal dada explícitamente en (2), cuyo ajuste a los datos del NO-2022 se muestran en la Fig. 3 y 4. Como podemos ver, la distribución (2) ajusta un poco menos al bulto de los resultados y en la cola sobre estima la cola. Así, el problema de colas sigue presente. Esto es atenuado puesto que la inmensa mayoría de los votos es compatible con una distribución Gamma.

Otra similitud importante entre el voto por el NO-2022 y el PRI-2026 es que ambos presentan votos en prácticamente todas las casillas. Es decir, hay una repulsión a no tener votos. Esto es especialmente importante para la dinámica pues indica que en casi todas las casillas de una elección nacional en un país megadiverso hay un agente que vota.

## IV. Discusión

Dado que las distribuciones aparecen como funciones teóricas bien establecidas, el siguiente paso es la búsqueda de mecanismos creíbles para tales resultados y una posible predicción de los parámetros de cada elección. La distribución Gamma es el resultado de un proceso de espera [16] sin embargo en el presente contexto no parece ser claro qué es lo que conduce a los resultados presentes. En esta sección discutimos dos modelos posibles sin dar números concretos.

Una opción, en términos de la dinámica de partido corporativo, es considerar los votos por casilla de acuerdo a cierta distribución y cuotas de votos por casilla o sección. En este modelo se considera que el partido cuenta con agentes en cada casilla y que requiere, además de sus copartidarios, convencer o coaccionar a cierto número de votantes. Se asume que los agentes conocen de antemano el número aproximado de votantes posible en otras casillas, por lo que destinan recursos para "convencer" votantes en donde tienen poco número de simpatizantes. Esto hace que se tenga consistentemente un número promedio y moda de votantes, no importando la elección. La existencia de agentes en todas las casillas asegura que no habrá casillas sin al menos un voto para el partido. Por supuesto que existen casillas con muchos afiliados por participar familiarmente con el partido. Las casillas están ordenadas de manera alfabética, de tal manera que, en zonas densamente pobladas habrá casilla con muchos familiares y hace razonable esta suposición. Este mecanismo fue explorado en la referencia [17] donde se presenta un mapeo de votos a distancias entre ciudades en un problema de Agente Viajero (TSP, por sus sigla en inglés). Un parámetro fundamental es la distribución inicial de ciudades. Se parte de una distribución de ciudades inicialmente en una cuadrícula, para la cual únicamente existen dos distancias posibles, y se inicia una relocalización de las ciudades como una probabilidad tomada de una gaussiana de centroide en el nodo de la cuadrícula $i$ y de varianza dada. Así el parámetro de la desviación standard lleva la distribución de ciudades de una con dos distancias permitidas a una que eventualmente tendrá las ciudades al azar y cuya distribución quasi optimal será

$$P_2(s) = N_2 x^{\wedge 2} \exp(-3s).$$
(3)

De nuevo con $N_2$, una constante de normalización.

En ese ejercicio se encontró que una solución compatible con una distribución Gamma ocurre cuando la distribución inicial de ciudades puede tener una posible superposición de ciudades al hacerse la desviación standard suficientemente grande. Traducido a votos, la idea es que los operadores políticos interactúen entre ellos conociendo el número de posibles votantes favorables por adelantado. La estrategia no es que en una casilla o unidad geográfica todos salgan a votar por su partido sino la de apoyar a las secciones más rezagadas en votos a alcanzar una cuota mínima. Es en este sentido un sistema interconectado que busca tener buenos resultados mínimos en la elección, que sean óptimos. El valor de los parámetros obtenidos en la referencia [17] no son de los que mejor ajustan a los resultados presidenciales, pero sí ofrecen una idea de cómo es que funciona el voto corporativo, pues recupera la idea de que existen cuotas de votantes en dichos partidos. Además recupera el fenómeno de colas anchas que se reporta en la Fig. 4.

Por otra parte, un modelo de agentes, en los cuales se mezclan agentes de cierta orientación de voto, con alguna distribución inicial y de ahí se tenga una bolsa de recursos para orientar el voto de otros ciudadanos, o la existencia de presiones en zonas ampliamente beneficiadas por los programas sociales del gobierno podría ser útil. En dichos modelos es usual que dé lugar a una ecuación tipo Fokker-Planck o similar. La distribución Gamma es la solución estacionaria de una ecuación de Fokker-Planck que tiene la forma

$$\partial_t f = (1/2)\,\partial_{ss}(s f) - (\beta/2)\,\partial_s((\alpha+1)/\beta - s).$$
(4)

Aquí el primer término del l.i.e. corresponde al término difusivo, mientras que el segundo a un término de arrastre. La solución estacionaria de (4) es justamente una distribución como la dada en (1). Por supuesto que aquí se requiere un modelo de agentes o de partículas que evolucione a la ecuación con los parámetros dados (véase [18], por ejemplo). Para ello el modelo tiene que ser calibrado y dicho modelado se encuentra en proceso [14].

## V. Conclusiones

Al autor le gustaría decir coloquialmente "No estoy diciendo que sea el PRI, pero es el PRI", sin embargo aún faltan aspectos para considerar, pues puede ser que sean las prácticas y el uso de recursos públicos no sean los únicos determinantes de la distribución de voto. El ejercicio revocatorio abre posibilidades por estar bien documentada y contener básicamente dos contribuciones al voto del NO: 1) los partidarios y núcleo duro del gobierno y 2) los votantes coaccionados para votar en cierto sentido. En los datos presentados aquí la distribución de voto es consistente no solo con el cuerpo principal, sino que la desviación en la cola de la distribución en ambos casos aparece. La ausencia de casillas con muy pocos votos es claramente el resultado de que pese a que la elección resultó de poco interés para muchos ciudadanos, prácticamente no hubo casillas con cero votos a favor del No. Es decir, se trata de un fenómeno nacional.

Por otra parte, modelos de tipo agente viajero o modelos de agentes en una red puede resultar útiles en la elucidación de los mecanismos de voto en México y en otros países. Sirva este texto para promover el estudio detallado de las elecciones y aprovechar situaciones particulares para aislar las dinámicas independientes en los procesos electorales.